\documentclass[showpacs, preprintnumbers,amsmath,amssymb,amsfonts,amsthm]{revtex4}

\usepackage{natbib, setspace, amsmath, amssymb, hyperref, epsfig, graphicx}

\begin{document}

\title{Earth's extensive entropy bound}

\author{Andreas Martin Lisewski}
\affiliation{
Baylor College of Medicine, One Baylor Plaza, Houston, TX 77030, USA
}%

\pacs{05.70.Ln, 95.30.Tg}


\begin{abstract}
{The possibility of planetary mass black hole production by crossing entropy limits is addressed. Such a possibility is given by pointing out that two geophysical quantities have comparable values: first,  Earth's total negative entropy flux integrated over geological time and, second, its extensive entropy bound, which follows as a tighter bound to the Bekenstein limit when entropy is an extensive function. The similarity between both numbers suggests that the formation of black holes from planets may be possible through a strong fluctuation toward thermodynamic equilibrium which results in gravothermal instability and final collapse. Briefly discussed are implications for the astronomical observation of low mass black holes and for Fermi's paradox.}
\end{abstract}
\bigskip

\maketitle

Any isolated and finite physical system will reach a state of thermodynamic equilibrium, where free energy has vanished and entropy has reached its highest possible level. For the known states of self-gravitating matter, such as main sequence stars, white dwarfs, or even neutron stars,  it can be estimated how long it would take these objects to reach thermodynamic equilibrium. The equilibrium point is characterized by a temperature balance with the surrounding space, which is a low temperature reservoir currently at $T \approx 3 $K.  For stellar objects, these time scales can be vast \cite{rees1997}: after consuming their thermonuclear fuel in approximately ten billion years, stars below the Chandrasekhar mass limit are expected to become white dwarf stars with lifetimes limited by the stability of protons, or at least  $10^{32}$ years.  
\medskip

Planets have lesser mass and do not carry such significant sources of energy. Their fate mainly depends on the companion stars and, ultimately, on the physical state of the Universe. Planet Earth today is a thermodynamic system far from equilibrium. Its current state is to a large part determined by incoming solar energy and by a negative flux of radiation entropy, which puts the system away from equilibrium by producing negentropy \cite{szargut2003}. Low entropy electromagnetic radiation from the Sun is absorbed, its free energy is successively dissipated in the Earth's atmosphere and surface, and higher entropy thermal radiation is released into space. Life and civilization, in their highly evolved and complex form, would be unlikely without this heat process \cite{schroedinger1944}. 
\medskip

The negative entropy flux $\dot\sigma$ can be estimated when Earth and Sun are approximated as black bodies. In this case, the Stefan-Boltzmann law gives \cite{kleidon2004, hermann2006}
\begin{equation}
\label{sbe}
\dot\sigma = I_0 (1-a) (1/T_{\rm E}-1/T_{\odot}) \approx 3.6 \,\rm{W \,m}^{-2}\rm{K}^{-1}\,,
\end{equation} 
where $a = 0.3 $ is the Earth's planetary albedo, $I_0 \approx 1368 \rm{\,W m}^{-2}$ is the net energy flux from the Sun to the Earth's surface, and $T_{\rm E} = 255$ K and $T_{\odot}=5760$ K are the black body radiation temperatures of Earth and Sun, respectively. Solar radiation energy is disspiated and entropy increased through processes such as absorption, heating and evaporation \cite{szargut2003}. In a steady state, $\dot\sigma$ balances the production of entropy $\dot\sigma_p$, i.e., $\dot\sigma \approx \dot\sigma_p$. However, the existence of non-equilibrium geological (fossils), atmospheric (chemical composition of the atmosphere), and biological (DNA replication, protein synthesis) structures and processes, which have been preserved over geological time $\lesssim \!\Delta \tau = 4 \times 10^9$ years, suggest that $\dot\sigma_p$ has been smaller than $\dot\sigma$. Therefore, Earth likely has gone through periods where negentropy $S$ has been accumulated in the system. This stored negentropy would be spontaneously released as additional entropy if the Earth would fail to support its peculiar non-equilibrium state, for example through a temporary excess in internal or external dissipative forces. Such destructive forces can likely be triggered by impacts of extraterrestrial massive objects (asteroids, comets) and radiation (supernovae, gamma-ray bursts), or internally by seismic activity (volcanos, earthquakes) and warfare (weapons of mass destruction). Those events have the potential to increase Earth's entropy load almost instantaneously through a high excess rate of entropy over negentropy production, i.e. $\dot \sigma_p \gg  \dot\sigma$.
\medskip

To establish an upper bound for $S$, it is assumed that Eq. (\ref{sbe}) is valid throughout geologi\-cal times, yielding $S \leq S_{\rm max}=\int \dot\sigma dA d\tau \approx \dot\sigma A_{\rm E}\Delta \tau/(k_{\rm B} \log 2) = 5 \times 10^{54}$ bits, where $k_{\rm B}$ is Boltzmann's constant and $A_{\rm E}=\pi R_{\rm E}^2$ is Earth's surface area facing the sun with radius $R_{\rm E}=6 \times10^6$ m. This assumption includes that radiation from the Sun was roughly constant during $\Delta \tau$, originating from thermonuclear burning of hydrogen in a main sequence star. The upper bound $S=S_{\rm max}$ represents an extreme situation where the entire negative entropy influx is stored, although, more realistically, dissipative processes will prevent $S$ to fully reach this limit. It is next shown that a comparable limit on $S$ can be imposed by considering gravitational entropy bounds.
\medskip

Bekenstein introduced a general entropy bound \cite{bekenstein1981} for any space region $\mathcal{O}$ which extends to a radius $R$ and contains energy $E$. He argued that if the total entropy $S$ associated with $\mathcal{O}$ is larger than the corresponding entropy $S_{\rm BH}$ of a black hole with energy $E_{\rm BH}>E$ and Schwarzschild radius $R_{\rm BH}>R$,
\begin{equation}
\label{bekb}
S_{\rm BH}=\frac{2 \pi E R}{\hbar \,c\,\log 2}\,,
\end{equation}  
then $\mathcal{O}$ itself cannot be a black hole, but can be made into one by an adiabatic process which slowly increases $E$ until a black hole is formed ($\hbar=h/2\pi$ is Planck's constant and $c$ is the vacuum speed of light.) Because the resulting object has entropy $S_{\rm BH}$ and entropy could not have been reduced during the adiabatic formation process, it follows that $S$ should be less or equal $S_{\rm BH}$ in the first place, and therefore $S < S_{\rm BH}$. 
\medskip

The Bekenstein bound can be further tightened if entropy is an extensive function of volume $V \!\sim R^3$ , which means that $S$ is homogeneous with $S(E,V) = V \, F(E/V)$ and with some function $F$.  In the present setting, the extensivity condition for entropy is directly ensured through the black body assumption which lead to Eq. (\ref{sbe}), i.e. $S \sim V \,T^3$.  Under this assumption the Bekenstein bound  necessarily follows a different scaling \cite{gour2003},
\begin{equation}
\label{ext}
S<S_{\rm ext}=S_{\rm BH}^{3/4},
\end{equation}
which results in a stricter bound if $S_{\rm BH}\gg1$. Eq. (\ref{ext}) represents the {\it extensive entropy bound}. To understand this result, it suffices to express the Bekenstein bound through dimensionless quantities,
\begin{equation}
\label{dimlb}
f(x)<\eta\,x\,y\,
\end{equation}
with 
\begin{equation}
\nonumber
\eta \equiv (6 \pi^2)^{1/3}, \quad x \equiv \frac{l_p^3E}{\epsilon_pV}, \quad y\equiv \frac{V^{1/3}}{l_p}, \quad f(x) \equiv l^3_p F(E/V)\,,
\end{equation}
and where $l_p$ and $\epsilon_p$ are the Planck length and Planck energy, respectively. A lower bound for $f(x)$ is obtained by minimizing $y$ as an independent variable, and a natural choice for a minimum scale at which energy is not lost significantly due to quantum fluctuations is the condition $R \sim \hbar/E$, or  $y \sim x^{-1/4}$. From this it follows that $f(x) < x^{3/4}$ and therefore the bound in Eq. (\ref{ext}). 
\medskip

Equation (\ref{ext}) represents the radiation entropy formula from the Stefan-Boltzmann law \cite{gour2003}, and it further suggests that thermodynamics of black holes and black bodies are physically related. The relationship can be made explicit for planet Earth as follows.
\begin{figure}[t]
\begin{center}
\epsfig{file=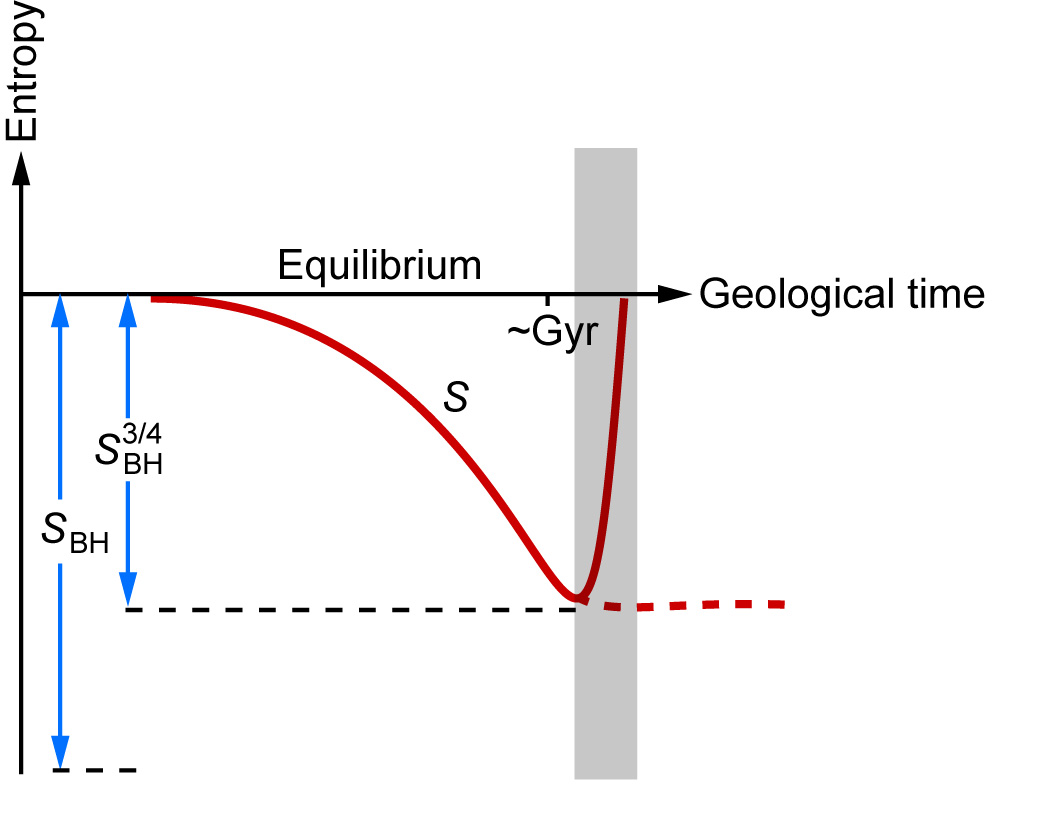, width=3.25in}
\end{center}
\caption{Schematic representation of a hypothetical increase in negentropy $S$ over geological time away from equilibrium. After $\Delta \tau \!\sim$ Gyr the total negentropy
$S$ becomes comparable to $S^{3/4}_{\rm BH}$. Upon loss
of this negentropy, entropy increases spontaneously and planetary black hole formation occurs as a result of a gravothermal instability (solid red line over grey box). In a second scenario, sustained accumulation of negentropy enters a steady state with entropy production and continues without large and potentially catastrophic fluctuations (dashed line).}
\end{figure}
With $R=R_{\rm E}$ and $M=M_{\rm E}=6\times 10^{27}\,$g it is $S_{\rm BH}=9\times10^{74}$ bits, a number that is 20 orders of magnitude larger than the  limit $S_{\rm max}=5\times10^{54} \,\rm{bits}$. This large difference reflects the orthodox view that for planets the Bekenstein bound is an unrealistic barrier. However, if it is reasonably assumed that Earth's entropy is an extensive function of volume, then Eq. (\ref{ext}) gives
\begin{equation}
\label{exergyearth}
S_{\rm ext}=2\times 10^{56} \,{\rm{bits}} \approx S_{\rm max} \,.
\end{equation}
Thus, Earth's  extensive entropy bound $S_{\rm ext}$ and maximal negentropy $S_{\rm max}$ are roughly within the same order of magnitude, suggesting that planetary and black hole physics may not be independent.
\medskip

Since no extensive system can have higher entropy than $S_{\rm ext}$, what would be the physical identity of an object after crossing the extensive entropy limit? One speculative scenario for the Earth is the spontaneous loss of accumulated negentropy (Fig. 1) leading to an immense increase of its internal kinetic energy by a total amount of  $E \approx k_{\rm B} \, T_{\rm E} \, S_{\rm  ext} \approx 7\times10^{35} \,{\rm J}$; and due to the virial theorem the total sum of gravitational potential and kinetic energy $E_t$ becomes  the negative of the total kinetic energy, or $E_t = -E$. At this point the planet will become unstable due to the gravothermal instability \cite{lyndenbellwood1968}: spherically contained  self-gravitating systems have {\it negative} heat capacity and this allows thermodynamically stable configurations between the inner core and the outer layers only if the system's radius is smaller than the critical value
\begin{equation}
\label{rc}
r_c = 0.335 \, \frac{G M^2}{-E_t} \,,
\end{equation}
but with $M=M_{\rm E}$ and $E = 7 \times 10^{35}$J it is $r_c \approx 1 {\rm km} \ll R_{\rm E}$ which clearly violates the stability condition. This also means that even if a fraction less than $10^{-3}$ of the extensive entropy bound would be stored as negentropy over geological times, a gravothermal instability would be reached. From here on the system  would condensate into a hot dense core surrounded by a less hot halo, whereupon the core would continue to collapse at an exponential rate while increasing its temperature faster than the halo. During central core-halo formation the system would also lose the extensivity property due to long range gravitational interactions. The likely end result of this catastrophic process would be a fall into a trapped surface and thus a formation of a central black hole \cite{glass2010}. 
\medskip

If similar physical conditions exist on other planets, then the formation of planetary black holes after reaching extensive entropy limits could result in testable astronomical predictions. Thus far, the main theoretical candidates for planetary mass ($M_{\rm BH}\lesssim 10^{30}$g) black holes have been primordial black holes \cite{carr2010}, which could have emerged from strong density fluctuations in the early Universe. Because the young Universe has cooled rapidly and strong density fluctuations have vanished, the total population of primordial black holes could not have increased ever since. Also, processes such as Hawking radiation and black hole mergers would lead to a smaller population. In contrast, Earth-like planets in stellar systems may reach their extensive entropy bounds and become black holes after gigayears of stellar evolution. This implies that the population of planetary black holes should increase during cosmological time. It is notable that the astronomical observation of planetary mass black holes may be within reach; for example, an appreciable density of black holes in the mass range $10^{20} {\rm g} < M_{\rm BH} < 10^{26}$g is currently not inconsistent with experimental results from microlensing and from constraints on gravitational waves \cite{carr2010}. 
\medskip

Another aspect of planetary black holes regards {\it Fermi's paradox.} This apparent paradox says that the vast spatial and temporal dimensions of the Universe seem incompatible with the absolute lack of evidence for extraterrestrial civilizations (``Where is everybody?''). Such lack of evidence may hint that planetary black holes form very efficiently before civilizations can escape their host planets through advanced space travel. Technological progress accompanied by industrial exploitation and global warfare may trigger those dissipative forces that lead to a gravothermal instability. Alternatively, an accelerating rate of extinction among biological species on Earth may be an indicator that entropy production has surpassed negentropy accumulation. How much time would a fluctuation towards equilibrium take?  For a sufficiently isolated thermodynamic system, the most probable path is along maximum entropy production and maximum entropy is reached within the system's relaxation time \cite{martyushev2006}; for the Earth's mantle, a relaxation time can be estimated by the ratio of viscosity to shear modulus, or around several hundred years. If this is the case, then intelligent life forms may exist on other Earth-like planets which deliberately keep technological progress, including development of interplanetary communication, at a slow pace, so to not perturb catastrophically their habitat's non-equilibrium state.     

\bibliography{entropy_lisewski_F}
\end{document}